\documentstyle[prl,aps,epsf]{revtex}
\begin{document}
\twocolumn[\hsize\textwidth\columnwidth\hsize\csname@twocolumnfalse%
\endcsname

\draft
 
\title{Spin Bose Glass Phase in Bilayer Quantum Hall Systems at $\nu=2$}
\author{Eugene Demler$^{1}$ and S. Das Sarma$^{1,2}$}
\address{
$^{1}$Institute for Theoretical Physics, University of California, Santa Barbara, CA 93106
}
\address{
$^{2}$Department of Physics, University of Maryland, College Park, Maryland 20742-4111
}
\date{\today}
\maketitle
\begin{abstract}
We develop an effective spin theory to describe magnetic properties of
the $\nu=2$ Quantum Hall bilayer systems.  In the absence of
disorder this theory gives quantitative agreement with the results of
microscopic Hartree-Fock calculations, and for finite disorder it
predicts the existence of a novel spin Bose glass phase. The Bose
glass is characterized by the presence of domains of canted
antiferromagnetic phase with zero average antiferromagnetic order and
short range mean antiferromagnetic correlations. It has
infinite antiferromagnetic transverse susceptibility, finite
longitudinal spin susceptibility and specific heat linear in
temperature. Transition from the canted antiferromagnet phase to the
spin Bose glass phase is characterized by a universal value of the
longitudinal spin conductance.
\end{abstract}


]

%\newpage

Recently a canted antiferromagnetic phase has
been predicted in bilayer quantum Hall (QH) systems at a total
filling factor $\nu=2$  on the basis of 
microscopic Hartree-Fock
calculations and a long wavelength quantum $O(3)$ nonlinear sigma model
\cite{DasSarma}. In this letter we construct
an alternative
effective spin theory that can describe the richness of the phase
diagram of a bilayer $\nu=2$ quantum Hall system. Our effective spin 
theory treats the interlayer tunneling nonperturbatively, in contrast to the
O(3) nonlinear sigma model which includes tunneling perturbatively
through an antiferromagnetic exchange. It gives 
excellent agreement with the results of microscopic Hartree-Fock
calculations in
\cite{DasSarma} and extends the earlier effective field theory
by allowing  to study quantitatively the effect of a finite
gate-voltage between the layers and calculate intersubband excitation 
energies. Our theory can easily incorporate the effects of  
disorder and we predict that for any non-zero disorder there
is a new $\nu=2$ spin Bose glass quantum Hall phase which may be visualized
as domains of canted antiferromagnetic phase surrounded by
domains of fully polarized ferromagnetic or spin singlet phases.
In this system the Bose glass phase we predict is quite novel,
and we elaborate in this Letter on the origin and the properties of this new QH 
glass phase. Related disorder induced 
spin phase has been discussed in a different setting in reference 
\cite{Senthil}.

In the absence of interlayer interaction each layer of the $\nu=2$ bilayer system 
would be
in a fully spin polarized ferromagnetic $\nu=1$ incompressible QH state
with spins in both layers pointing in the direction of the applied
magnetic field (FPF state). Tunneling
between the layers favors the formation of spin singlet states from
the pairs of electrons in the opposite layers and energetically stabilizes the spin singlet (SS) 
state. In \cite{DasSarma}
it was observed that the competition between the two tendencies may 
lead to a third intermediate phase: canted antiferromagnetic state, where spins in the two layers
have the same component along the applied field but opposite
components in the perpendicular 2D plane (CAF state).

\begin{figure*}[h]
\centerline{\epsfxsize=4cm 
\epsfbox{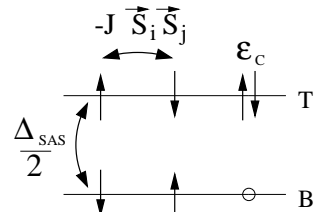}
}
\caption { Effective bilayer lattice model for $\nu=2$ double layer QH system }
\label{figure1}
\end{figure*}

We now introduce a simple lattice model which we use to describe the
physics of the bilayer $\nu=2$ QH system. We consider a bilayer lattice 
model shown in figure \ref{figure1}. 

Sites in each layer may be thought of as labeling different
intra-Landau-level states. Electrons may tunnel from one layer to
another conserving the in-plane
site index (i.e. between the states with the same intra-Landau-level index). There is
a ferromagnetic interaction between nearest-neighbor sites within
individual layers and a Zeeman interaction with the
applied magnetic field. 
We also account for the charging energy, i.e. the energy
cost of creating charge imbalance between the layers through the term ${\cal H}_c$ below. 
The Hamiltonian of the system may be written as

\begin{eqnarray}
{\cal H} &=& {\cal H}_{T} + {\cal H}_{c} + {\cal H}_{Z} + {\cal H}_{F } \nonumber\\
{\cal H}_{T} &=& - \frac{1}{2}\Delta_{SAS} \sum_i ( c_{Ti\sigma}^{\dagger}
c_{Bi\sigma} + c_{Bi\sigma}^{\dagger} c_{Ti\sigma} )  \nonumber\\
{\cal H}_{c} &=& \frac{1}{2} \epsilon_c
\sum_i \left( (n_{Ti}-1)^2 + (n_{Bi}-1)^2 \right) \nonumber\\
{\cal H}_{Z} &=& -H_z \sum_i ( S^z_{Ti} + S^z_{Bi} ) \nonumber\\
{\cal H}_F &=&- J \sum_{\langle ij \rangle} ( {\bf S}_{Ti} {\bf S}_{Tj}
+ {\bf S}_{Bi} {\bf S}_{Bj} )
\label{Hamiltonian}
\end{eqnarray}
where $T$ or $B$ is the isospin index that labels electrons in the top and bottom layers
respectively, $i$ is the in-plane site ( intra-Landau-level ) index, 
and $\sigma$ is the spin index.
${S}^a_{Ti} = c_{Ti\alpha}^{\dagger}
{\bf \sigma^a_{\alpha\beta}} c_{Ti\beta}$ and $n_{Ti} = \sum_{\sigma} c_{Ti\sigma}^{\dagger}
c_{Ti\sigma}$ are spin and charge operators for layer $T$, 
with analogous definitions for layer $B$.
Parameters $J$ and $\epsilon_c$ of this model may be easily estimated
as  $\rho_S^0=e^2/(16\sqrt{2 \pi} \epsilon l)$ ,
where $l=\sqrt{\hbar c/e B}$ is magnetic length and 
$\epsilon_c =
\frac{e^2}{\epsilon l}\sqrt{\frac{\pi}{2}}( 1
-e^{-d^2/l^2}Erf[d/(l\sqrt{2})])$, where $d$ is the distance between 
the layers and $Erf$ is the error function \cite{long}.

Let us consider an individual rung, i.e. two sites with the same in-plane site 
index on the opposite layers.  Each rung must be populated by two electrons, therefore we
have six possible states for each rung. 
The six available states are conveniently
classified into three states that are spin triplets 
\begin{eqnarray}
\begin{array}{ll}
| t_{+} \rangle = t_{+}^{\dagger} | 0 \rangle &= c_{T\uparrow}^{\dagger} c_{B\uparrow}^{\dagger} | 0 \rangle
\\ 
 | t_{0} \rangle = t_{0}^{\dagger} | 0 \rangle &= \frac{1}{\sqrt{2}} \left(
c_{T\uparrow}^{\dagger} c_{B\downarrow}^{\dagger}
+ c_{T\downarrow}^{\dagger} c_{B\uparrow}^{\dagger} \right)
 | 0 \rangle
\\ 
| t_{-} \rangle = t_{-}^{\dagger} | 0 \rangle &= c_{T\downarrow}^{\dagger} c_{B\downarrow}^{\dagger} | 0 \rangle
\end{array} 
\label{tstates}
\end{eqnarray}
and three states that are spin singlets 
\begin{eqnarray}
\begin{array}{ll}
| \tau_{+} \rangle = \tau_{+}^{\dagger} | 0 \rangle &= c_{T\uparrow}^{\dagger} c_{T\downarrow}^{\dagger} | 0 \rangle
\\ 
 | \tau_{0} \rangle = \tau_{0}^{\dagger} | 0 \rangle &= \frac{1}{\sqrt{2}} \left(
c_{T\uparrow}^{\dagger} c_{B\downarrow}^{\dagger}
- c_{T\downarrow}^{\dagger} c_{B\uparrow}^{\dagger} \right)
 | 0 \rangle
\\ 
| \tau_{-} \rangle = \tau_{-}^{\dagger} | 0 \rangle &= c_{B\uparrow}^{\dagger} c_{B\downarrow}^{\dagger} | 0 \rangle
\end{array}
 \label{taustates}
\end{eqnarray}
Operators  $t$ and $\tau$ satisfy bosonic commutation relations \cite{Sachdev}
and constraint 
$
\tau_{\alpha}^{\dagger} \tau_{\alpha} + t_{\alpha}^{\dagger} t_{\alpha} =1
$
projects into the physical Hilbert space.

In Hamiltonian (\ref{Hamiltonian}) all the terms except ${\cal H}_F$ act within a single rung.
It is therefore  natural as a first step to diagonalize ${\cal H}' = {\cal H}_{T} + 
{\cal H}_{c} + {\cal H}_{Z}$ on one rung. The latter task is simplified by the observation that
${\cal H}_{T}$ and ${\cal H}_{c}$ act in the subspace of $\tau$ states, whereas
${\cal H}_{Z}$  operates in the $t$ subspace. A simple calculation gives for the lowest energy
eigenstates of  ${\cal H}'$: 
\begin{itemize}
\item state $| t_{+} \rangle $  with energy $ E_t = - H_z $ 
\item state 
$
| v_{+} \rangle = \frac{ ( \sin\theta - \cos \theta ) }{2}  
( | \tau_{+} \rangle + | \tau_{-} \rangle ) 
-\frac{ ( \sin\theta + \cos \theta ) }{\sqrt{2}} | \tau_{0} \rangle 
$ \newline
with  energy $E_v = \frac{\epsilon_c}{2} - \sqrt{\Delta_{SAS}^2 + \epsilon_c^2/4}$. Here
$ \tan \theta = \epsilon_c/( 2 \Delta_{SAS} +  2 \sqrt{\Delta_{SAS}^2 + \epsilon_c^2/4} )
$
\end{itemize}
State $| v_{+} \rangle $ is a spin-singlet state whose energy is lowered by
interlayer tunneling and $| t_{+} \rangle $ is a spin triplet state favored by Zeeman interaction.
Competition between the two states is a competition between the SS state 
and the FPF state. In the absence of the in-plane ferromagnetic interaction
we would have level crossing at $E_v = E_t $ with a first
order phase transition between SS and FPF phases. However as we show below  ${\cal H}_{F}$
acts as an interaction that connects the two states and gives rise to an intermediate
state that is a superposition of the $| v_{+} \rangle $ and $| t_{+} \rangle $ states and 
corresponds to the CAF phase. 

We rewrite Hamiltonian (\ref{Hamiltonian}) keeping only the lowest energy states 
$| v_{i+} \rangle $ and $| t_{i+} \rangle $:
\begin{eqnarray}
&&\tilde{{\cal H}} = E_t \sum_i t_{i+}^{\dagger}  t_{i+} + E_v  \sum_i v_{i+}^{\dagger}  v_{i+}
\nonumber\\
&-& \frac{J}{4} ( \cos \theta + \sin \theta )^2 \sum_{\langle ij \rangle}
\left( t_{i+}^{\dagger} v_ {i+}v_{j+}^{\dagger} t_{j+} 
+ t_{j+}^{\dagger} v_{j+} v_{i+}^{\dagger} t_{i+}
\right) \nonumber\\
&-& \frac{J}{2}  \sum_{\langle ij \rangle} t_{i+}^{\dagger}  t_{i+} t_{j+}^{\dagger} t_{j+}
\label{Hred}
\end{eqnarray}
and the hard core constraint is implied
\begin{eqnarray}
v_{i+}^{\dagger} v_{i+} + t_{i+}^{\dagger}  t_{i+} = 1
\label{constraint1}
\end{eqnarray}

\begin{figure*}[h]
\centerline{\epsfxsize=6cm 
\epsfbox{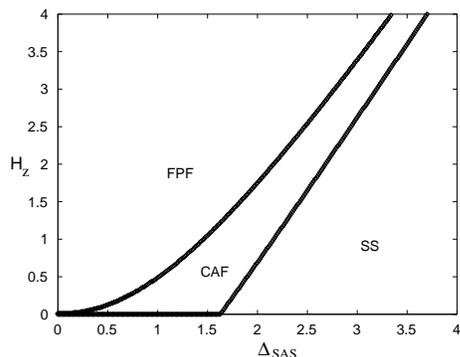}
}
\caption { Phase diagram for disorder-free system. All energies are in units of $J$, $\epsilon_c =5.0 J$. }
\label{figure2}
\end{figure*}

The mean field
analysis of Hamiltonian (\ref{Hred}) may be done by considering states with simultaneously
condensed $v$ and $t$ bosons. They correspond to the variational wavefunctions
of the form 
\mbox{ $
| \Psi \rangle = exp\{ \alpha \sum_i v_{i+}^{\dagger} + \beta  \sum_i t_{i+}^{\dagger}\} | 0 \rangle
$} \cite{Sachdev}.
The energy of state $| \Psi \rangle$ is given by
\begin{eqnarray}
E_0 = 
E_v | \alpha |^2 + E_t | \beta |^2 - J ( \cos \theta + \sin \theta )^2 |\alpha|^2 |\beta|^2 
- J |\beta|^4
\label{Eab}
\end{eqnarray}
and state $| \Psi \rangle $ obeys constraint (\ref{constraint1}) on the average provided that
\begin{eqnarray}
| \alpha |^2 + | \beta |^2 =1
\label{constraint2}
\end{eqnarray}
Values of $\alpha$ and $\beta$ that minimize (\ref{Eab}) under the condition 
(\ref{constraint2}) are given by
\begin{eqnarray}
\begin{array}{lll}
| \alpha | =1 & | \beta | = 0 & if~t_{min}<0 \\
| \alpha | = t_{min} & | \beta | = \sqrt{ 1 - t_{min}^2} & if~0 < t_{min} < 1 \\
| \alpha | =0 & | \beta | = 1 & if~t_{min}>1
\end{array}
\label{ab-phases}
\end{eqnarray}
where
\begin{eqnarray}
t_{min} = \frac{ J ( \cos \theta + \sin \theta)^2 - 
E_t + E_v}{2 [ J ( \cos \theta + \sin \theta)^2 - J]}
\label{tmin}
\end{eqnarray}
The first and the last cases obviously correspond to the FPF and SS phases respectively 
\cite{comment}.
But there is also a nontrivial new phase that appears in our analysis when both 
$|\alpha |$ and $|\beta|$ are finite. It is easy to verify that this state corresponds 
precisely to the canted
antiferromagnetic phase discussed in \cite{DasSarma}
with direction of the Neel ordering given by the phase between the $t_+$ and $v_+$ 
condensates 
$ tan^{-1} N_y/N_x = Arg ( \alpha^* \beta )$ ( Neel order parameter is defined as 
 ${\bf N} = \sum_i {\bf S}_{Ti} - \sum_i {\bf S}_{Bi}$ ).
On figure \ref{figure2} we show the phase diagram obtained from equation (\ref{ab-phases}).
We can also use our bosonic model to calculate the phase diagram in the presence of an
interlayer charge imbalance and these results will be reported elsewhere \cite{long}.

The lowest energy interband transition in the SS phase will correspond to destroying
a $v_+$ and creating a $t_+$  boson. The energy for such transition is
$
\omega_- = - J ( cos \theta + sin \theta )^2 + E_t - E_v
$
and vanishes at the SS/CAF transition as may be seen from equations 
(\ref{ab-phases}) and 
(\ref{tmin}).  Analogously in the FPF state the lowest energy interband transition
will correspond to destroying $t_+$ and creating a $v_+$ boson \cite{long}.

\begin{figure*}[h]
\centerline{\epsfxsize=8cm 
\epsfbox{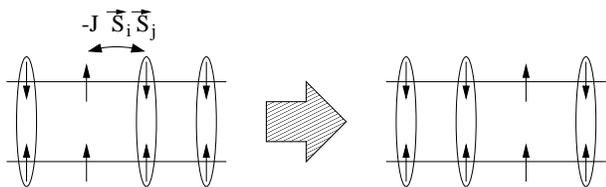}
}
\caption { Triplet ``hopping'' process due to ferromagnetic in-pane interaction }
\label{figure3}
\end{figure*}

Let us now give a simple physical picture that will illustrate formal
calculations presented above. We consider an SS state which has
singlet $v_+$-bosons on all rungs and imagine creating $t_+$-triplet on one of
the rungs.  Creating a localized triplet requires energy $ E_t-E_v $ and 
this energy is
unaffected by the ferromagnetic interaction since
parallel and anti-parallel contributions cancel for triplet
interacting with neighboring singlets. 
However ${\cal H}_F$ also gives rise to 
a process in which one of the spins of the triplet pair and one
spin from the neighboring singlet pair are flipped
simultaneously. This process is shown on figure \ref{figure3} and may be
interpreted as hopping of the triplet boson to the nearest-neighbor
site. 
Therefore creating a propagating
triplet boson at wavevector $k$ will give it an additional kinetic energy
$\tilde{J} ( cos k_x + cos k_y )$ due to ${\cal H}_F$. This allows us to
have a situation when $ E_t-E_v > 0 $ but $ E_t-E_v - 2 \tilde{J} < 0 $,
i.e. when it is energetically unfovarable to create localized
triplets but it is already favorable to create them at $k=0$, i.e. to have
a condensate of $t_+$ bosons. This effect is
the origin of the CAF state and allows us to understand this phase as a coherent
superposition of condensed $t_+$ and $v_+$ bosons. 

\begin{figure*}[h]
\centerline{\epsfxsize=6cm 
\epsfbox{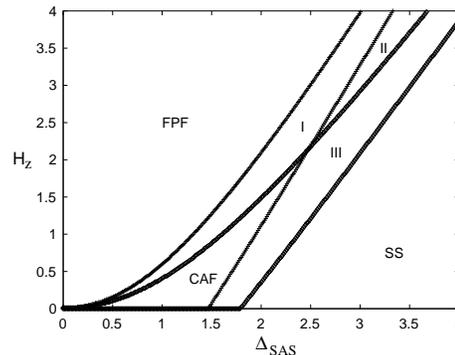}
}
\caption { Phase diagram for a disordered system. All energies are in units of $J$, $\epsilon_c = 5~J$.
Variation in $\Delta_{SAS}$ was assumed to be 10 \%.  Region I corresponds to domains
of CAF and FPF phases, Region III to domains of SS and CAF and region II to domain 
of all three kinds.
There are no phase transitions among regions I, II, III, they all correspond to 
the spin Bose glass phase. }
\label{figure4}
\end{figure*}

In a real system there is always disorder. It may be due to
fluctuations in the distance between the wells or the presence of
impurities. Such disorder can be easily included in our effective bosonic
theory, but would be difficult if not impossible, to include in the Hartree-Fock 
theory of references \cite{DasSarma}.
For our effective spin model the major effects of disorder will
be randomness in the value of tunneling $\Delta_{SAS}$ and the
appearance of a random local gate voltage, in both cases
leading to random local fluctuations in the energy of the $v_+$ boson.
Then if we are close to the CAF-SS transition we may have a situation induced by disorder where
$ E_t-E^{max}_v - 2 \tilde{J} < 0 $ and $E_t-E^{min}_v - 2 \tilde{J} > 0 $. So
for some regions creating
non-local $t_+$ triplets will lower the energy of the system
and for some regions it will lead to an energy increase. In this case the system breaks
into domains, with each domain being locally a CAF phase or a SS phase (region III on figure 
\ref{figure4}). 
Each CAF domain may be thought of as being in a quantum disordered state
with an undefined direction of the Neel order but finite z-magnetization \cite{Senthil}.
Close to the 
CAF-FPF transition line in the disorder-free system we may have  
a disorder induced situation where we have  CAF domains
in the background of domains of the FPF phase (region I on figure \ref{figure4}). 
Finally we can also have the phase where we have domains of all three kinds 
( region II on figure \ref{figure4} ).
In figure \ref{figure4} we show the resulting phase diagram for the same values of parameters as in figure \ref{figure2} but assuming that $\Delta_{SAS}$
may randomly vary by 10\% around its average value. Such a variation in $\Delta_{SAS}$ is 
physically
quite reasonable even in high quality 2D systems since $\Delta_{SAS}$ depends exponentially 
on layer thickness.
There are no phase transitions between regions I, II and III 
on the phase diagram in figure \ref{figure4} but only smooth crossovers.
The  true quantum phase transitions occur between FPF and I, SS and III and between CAF and one 
of the I, II or III regions.

The nature of these phase transitions is also easy to understand.
The SS phase is
an insulating phase of zero density of  $t_{+}$ bosons, FPF state is an insulating phase
with density $n=1$\cite{comment} and CAF is a superfluid phase.
Randomness that we consider acts as a randomness in the chemical potential of these $t_{+}$ bosons,
so our problem is equivalent to the problem of bosons in a random potential, the so-called dirty boson problem, considered in references \cite{Fisher,Cha}. 
We immediately recognize I, II and III as a single Bose glass (SBG) phase
of the singlet and triplet bosons. This observation allows us to draw several important conclusions
about the properties of this SBG phase. 
In the SS state the $\langle S^z(-\omega) S^z(\omega) \rangle$ correlation
function is zero and in the CAF phase it has a $\delta$-function peak at zero frequency
due to the Goldstone mode of
the spontaneous breaking of the $U(1)$ symmetry of spin rotations around the $z$-axis.
In the SBG phase this correlation function
will be finite at small frequencies, which implies finite longitudinal spin susceptibility and  
is the analog of finite compressibility of the usual charge Bose glass. 
Our new SBG phase does not have antiferromagnetic long range order, i.e. 
$ \langle N_{x(y}) \rangle = 0 $.
All the $t_+$ and $v_+$ bosons are localized in this phase, therefore it will have 
only short range mean antiferromagnetic correlations. But 
analogous to the infinite superfluid susceptibility of charge Bose glass our SBG phase
will have an infinite transverse antiferromagnetic susceptibility.
Another important feature of the Bose glass phase is a finite density of low energy
excitations \cite{Fisher,Cha}. This implies that our SBG phase will have a specific heat linear in temperature which provides another way to experimentally distinguishing it 
from the CAF phase whose specific heat goes as $T^2$ or FPF and SS phases that have exponentially small specific heat at low temperatures.
The existence of the SBG
phase separating SS, FPF and CAF phases also has important consequences in that it changes
the critical exponents for the corresponding phase transitions from the one obtained in {\cite{DasSarma} 
for the disorder-free system. The new critical exponents will be those of the superconductor-insulator
transition in dirty boson system studied in \cite{Fisher,Cha}. We would also like to point out that
the SBG system that we suggested may be a better experimental realization of a 2d 
superconductor-insulator transition in a boson system than conventionally used 2d superconducting 
films \cite{Goldman} in that it is free of long range forces and allows one to vary the density of bosons 
by varying $H_z$. In addition our predicted QH Bose glass phase transition does not
have the complication  arising from parallel fermionic excitations which may play
a role in the superconducting films \cite{Hagenblast}. We therefore expect, based on 
the arguments given in \cite{Fisher,Cha}, that transition from the CAF phase to the 
SBG phase will be characterized by a truly universal longitudinal spin conductance, which in 
principle can be measured by measuring 
the spin susceptibility and the spin diffusion coefficient.

Before concluding we remark on the feasible experimental observability of our proposed
QH Bose glass phase. First we remark that the basic $\nu=2$ QH phase transition and the associated
softening of the relevant spin density excitations has been verified experimentally
\cite{Pellegrini} via inelastic light scattering spectroscopy. Since interlayer tunneling
fluctuations are invariably present in real systems, it is in fact quite possible that 
the experiments in \cite{Pellegrini} have already observed a transition to the Bose glass phase
as in our figure \ref{figure4}. Some evidence supporting this possibility comes from the fact that softening
of the spin density excitations observed in \cite{Pellegrini} did not lead to the appearance of a sharp
dispersing Goldstone mode expected in the CAF phase but only to some 
broad zero energy spectral weight consistent with the Bose glass phase. Future experiments
in samples with deliberately controlled disorder
should be carried out to conclusively verify our prediction of a QH disordered Bose glass phase.

In conclusion we predict a new 2D Bose glass phase in a $\nu=2$ QH bilayer system by
introducing an effective spin theory. This phase has the usual properties of a Bose glass
phase \cite{Fisher,Cha} including a universal spin conductance at the transition. While we
have specifically considered the $\nu=2$ integer QH situation, our arguments should
go through for all $\nu=2/$(odd~integer) fractional QH states also, following the reasoning of 
\cite{DasSarma}, and for the fractional filling there should be an exotic fractional quantum 2D Bose glass in bilayer systems\cite{disorder}.

This work is supported by the NSF at ITP and by the US-ONR (S.D.S.). 
We acknowledge useful discussions with M.P.A. Fisher, Y. Kim, A. MacDonald, 
R. Rajaraman, S. Sachdev, and T. Senthil.


\begin{thebibliography}{10}

\bibitem{DasSarma}
\newblock L. Zheng {\it et. al. }, {\em Phys. Rev. Lett.}, {\bf 78}:2453 (1997);
\newblock S. Das Sarma {\it et. al. }, {\em Phys. Rev. Lett.}, {\bf 79}:917 (1997);
\newblock  S. Das Sarma {\it et. al. }, {\em Phys. Rev. B} {\bf 58}:4672 (1998)

%\bibitem{Senthil-thesis}
%T. Senthil, Ph.D. Thesis, Yale University, unpublished
\bibitem{Senthil}
\newblock T. Senthil and S. Sachdev, {\em Annals of Physics} {\bf 251}:76 (1996)


\bibitem{long}
\newblock E. Demler and S. Das Sarma, unpublished

\bibitem{Sachdev}
\newblock S. Sachdev and R. Bhatt, {\em Phys. Rev. B} {\bf 41}:9323 (1990);
%\newblock  S. Goplan {\it et. al. }, {\em Phys. Rev. B} {\bf 49}:8901 (1994)




\bibitem{comment}
\newblock  In our variational wavefunctions $ | \Psi \rangle$
the FPF phase is described as a superfluid state of $t_+$ bosons with 
average density $n=1$ rather than an insulating state with exactly
one $t_+$-boson per site. This is a
result of our mean-field treatment of the hard core constraint (\ref{constraint2}).
Analogously the SS state in our mean-field calculations appears as a superfluid of $v_+$ bosons
with one boson per site on the average
and not as an insulator with exactly one $v_+$ boson per site.



\bibitem{Fisher}
\newblock  M.P.A. Fisher {\it et. al. }, {\em Phys. Rev. B} {\bf 40}:546 (1989);
\newblock  M.P.A. Fisher {\it et. al. }, {\em Phys. Rev. Lett.}, {\bf 64}:587 (1990)

\bibitem{Cha}
\newblock M. Cha  {\it et. al. }, {\em Phys. Rev. B} {\bf 44}:6883 (1991);
\newblock M. Wallin  {\it et. al. }, {\em Phys. Rev. B} {\bf 49}:12 115 (1994)



\bibitem{Goldman}
\newblock A. Hebard and M. Paalanen,  {\em Phys. Rev. Lett.} {\bf 65}:587 (1990);
\newblock Y. Liu {\it et. al. },  {\em Phys. Rev. B} {\bf 47}:5931 (1993);
\newblock A. Yazdani and A. Kapitulnik,  {\em Phys. Rev. Lett.} {\bf 74}:3037 (1995)

\bibitem{Hagenblast} 
\newblock K. Hagenblast {\it et. al. },  {\em Phys. Rev. Lett.} {\bf 78}:1779 (1997);

\bibitem{Pellegrini}
\newblock  V. Pellegrini {\it et. al. },  {\em Phys. Rev. Lett.} {\bf 79}:310 (1997);
\newblock  V. Pellegrini {\it et. al. },  {\em Science} {\bf 281}:799 (1998)

\bibitem{disorder}
Another plausible effect of disorder, which we do not discuss in this paper,
is the appearance of unmatched spins $1/2$ leading to the possibility
of a random-singlet phase. This effect is not important at $\nu=2$ but may
become relevant for the  $\nu=2/$(odd~integer) fractional QH states.



\end{thebibliography}
\end{document}